\documentclass[aps,pre,showpacs,showkeys]{revtex4}
\begin{document}
%
%
\title{ Generalized  kinetic and evolution equations  in the approach of the 
nonequilibrium  statistical operator }

%
%
\author{A.\ L.\ Kuzemsky }
\affiliation{Bogoliubov Laboratory of Theoretical Physics,\\
 Joint Institute for Nuclear Research, 141980 Dubna, Moscow Region, Russia}
\email{kuzemsky@thsun1.jinr.ru}
\date{\today}
%
%
%
\begin{abstract}
%
The method of the nonequilibrium statistical operator developed by D. N. Zubarev is
employed to analyse and derive   generalized transport and kinetic equations. 
The degrees of freedom
in solids   can  often be represented  as  a few interacting subsystems (electrons, spins, phonons, nuclear spins,
etc.).  Perturbation of one subsystem may produce a nonequilibrium state which is then  relaxed
to an equilibrium state due to the interaction between  particles or with a thermal bath.
The generalized kinetic equations  were derived for a system weakly coupled to a thermal bath
to elucidate the nature of transport and relaxation processes.  It was shown that the
"collision term" had the same functional form as for the generalized kinetic
equations for the system with small interactions among  particles.
The applicability of the general formalism to physically relevant situations is investigated.
It is shown that some 
known generalized  kinetic equations (e.g. kinetic equation for  magnons,
Peierls equation for phonons)  naturally  emerges  within the NSO formalism. 
The relaxation of a small dynamic  subsystem in contact with a thermal bath is considered on the basis of the
derived equations.
The Schrodinger-type equation for the average amplitude  describing the energy shift and damping of a
particle in a thermal bath  and the coupled kinetic equation  describing the dynamic  and statistical
aspects of the motion  are derived and analysed. The equations derived  can help in the understanding of the origin of
irreversible behavior in quantum phenomena.  
%
\end{abstract}
%
\maketitle
%
%
%
\section{Introduction}
%
%
%
%
The aim of statistical mechanics is to give a consistent
formalism for a microscopic description of   macroscopic behavior of matter 
in bulk.
The methods of equilibrium and nonequilibrium 
statistical mechanics have been fruitfully applied to a large
variety of phenomena and materials~\cite{bb,cha88,grand87,grand88,toda92,kubo91,donald00}.
The statistical mechanics of irreversible processes in solids, liquids, and complex
materials like a soft matter are at the present time of much interest. 
The central problem of nonequilibrium statistical mechanics is to derive a set of equations which
describe irreversible processes from the reversible equations of motion.
The consistent calculation of transport coefficients
is of particular interest because  one can get information on the microscopic
structure of the condensed matter. There exist a lot of theoretical methods for  calculation
of transport coefficients as a rule having a fairly restricted range of validity and
applicability.  The most extensively developed theory of transport processes is that based on the
Boltzmann equation~\cite{boltz,pei74}. However, this approach has strong restrictions and can reasonably be applied 
to a strongly rarefied gas of point particles~\cite{uh74}. For systems in the state of statistical equilibrium, there
is the Gibbs distribution~\cite{gibbs} by means of which it is possible to calculate an average value 
of any dynamical quantity. No such universal distribution has been formulated for irreversible
processes. Thus, to proceed to the solution of problems of statistical mechanics of
nonequilibrium systems, it is necessary to resort to various approximate 
methods~\cite{green54,hgreen54,mont54,mori56,mori58,hmori58,mori59}. Kubo and 
others~\cite{kubo57,rkubo57,sadao58}
 derived the quantum statistical expressions for transport coefficients such as electric and
thermal conductivities. They considered the case of mechanical disturbances such as an electric field.
The mechanical disturbance is expressed as a definite perturbing Hamiltonian and the deviation from
equilibrium caused by it can be obtained by perturbation theory. On the other hand, thermal
disturbances such as density and temperature gradients cannot be expressed as a perturbing Hamiltonian
in an unambiguous way.
During the last decades, a number of schemes have been 
concerned with a more general and consistent approach to transport 
theory~\cite{zub61,prig62,mori65,hmori65,lee,zub74,grand88,macl89,ichi94,zwan01}.
These approaches, each in its own way, lead us to substantial advances in the understanding of the
nonequilibrium behavior of many-particle classical and quantum systems. In addition, they have used dynamic 
arguments to obtain kinetic  and balance equations which describe the irreversible evolution of a system from
particular initial states.  This field is very active and
there are many aspects to the problem~\cite{leb99}. 
The purpose of the present work is to elucidate further the nature of  transport processes and
irreversible phenomena from a dynamic point of view. According to Montroll~\cite{mont}, "dynamics is the science of cleverly
applying the operator $\exp (-iHt / \hbar)$".  
We wish to give a self-contained consideration of some general approach to the description of 
transport phenomena   starting with dynamic equations. 
Our purpose here is to discuss the derivation, within the formalism of the nonequilibrium
statistical operator~\cite{zub61,zub70,zub74}, of the generalized transport and kinetic equations. On this basis
we shall derive, by statistical mechanics methods, the kinetic equations for a system 
weakly coupled to a thermal bath.\\
In section \ref{nso}, we  briefly review some basic concepts. In section \ref{nsok}, the derivation of the
transport and kinetic equations within the NSO formalism is outlined. In section \ref{nsom}, we 
consider the application of the established equations to the derivation of the kinetic equations for
magnons and phonons. Special attention is given to the problem of derivation of kinetic equations for a system 
weakly coupled to a thermal bath in section \ref{nsob}. On the  basis of these equations the balance and 
master equations are obtained in section  \ref{nsobm}. The behavior of a small dynamic system weakly coupled to a thermal
bath is discussed in some detail in section  \ref{tb}. The relaxation of a small dynamic subsystem in contact with 
a thermal bath is considered on the basis of the derived equations.
The Schrodinger-type equation for an average amplitude describing the energy shift and damping of a
particle in a thermal bath, and the coupled kinetic equation  describing the dynamic and statistical
aspects of the motion  are derived and analysed in section \ref{seq}. 
%

%
%
\section{Outline of the nonequilibrium statistical operator method}\label{nso}
%
%
In this section, we  briefly   recapitulate 
the main ideas of the  nonequilibrium statistical operator approach~\cite{zub61,zub70,zub74} for the sake of a self-contained 
formulation. The central statement of the statistical-mechanical picture is the fact that it is practically 
impossible to give a complete description of the state of a complex macroscopic system. We must substantially  reduce 
 the number of variables and confine ourselves to the description of the system which is considerably
less then complete. The problem of predicting  probable behavior of a system at some specified 
time is a statistical one. As it was shown by Gibbs~\cite{gibbs} and Boltzmann~\cite{boltz}, it is useful and
workable to employ the technique of representing the system by means of an ensemble consisting of a 
large number of identical copies of a single system under consideration. The state of the ensemble
is then described by a distribution 
function $\rho (\vec r_{1} \ldots \vec r_{n}, \vec p_{1} \ldots \vec p_{n}, t)$ in the phase space
of a single system. This distribution function is  chosen so   that averages over the ensemble are in exact 
agreement with the incomplete ( macroscopic ) knowledge of the state of the system at some specified time.
Then the expected development of the system at subsequent times is modelled via the average behavior of members of the
representative ensemble. It is evident that there are many different ways in which an ensemble could be
constructed. As a result, the basic notion, the distribution function $\rho$ is not uniquely defined.
Moreover, contrary to the description of a system in the state of thermodynamic equilibrium which is
only one for fixed values of volume, energy, particle number, etc., the number of 
nonequilibrium states is large. The role of the relaxation times to equilibrium state was analysed in paper~\cite{gal}.
The precise definition of the nonequilibrium state is quite difficult
and complicated, and is not uniquely specified. Since it is virtually impossible and impractical to
try to describe in detail  the state of a complex macroscopic system in the nonequilibrium state, the
method of reducing the number of relevant variables was widely used. A large and important class
of transport processes can  reasonably be  modelled in terms of a reduced number of macroscopic
relevant variables~\cite{lew67}. There are different time scales and different sets of the relevant variables~\cite{bog,bog1},
e.g. hydrodynamic, kinetic, etc. This line of reasoning has led to  seminal ideas on the
construction of   Gibbs-type ensembles for  nonequilibrium systems~\cite{berg55,leb57,leb62,macl89}.
B. Robertson~\cite{rob66,rob67,rob70,rob71} proposed the method of equations of motion for the "relevant" variables, the
space- and time-dependent thermodynamic "coordinates" of a many-body nonequilibrium system  which were derived
directly from the Liouville equation. This was done by defining a generalized canonical density
operator depending only upon present values of the thermodynamic "coordinates".
The most satisfactory and workable approach 
to the construction of  Gibbs-type ensembles for the nonequilibrium systems, as it appears to the writer, is the method of 
nonequilibrium statistical operator (NSO) developed by D. N. Zubarev~\cite{zub74}. The NSO method
permits one to generalize the Gibbs ensemble method~\cite{gibbs} to the nonequilibrium case and to
construct a nonequilibrium statistical operator which  enables one to obtain the transport equations and
calculate the kinetic coefficients in terms of correlation functions, and which, in the case of
equilibrium, goes over to the Gibbs distribution. Although this method is well known, we shall
briefly recall it, mostly in order to introduce the notation needed in the following.\\
The NSO method sets out  as follows. The irreversible processes which can be considered as a reaction
of a system on mechanical perturbations can be analysed by means of the method of linear reaction
on the external perturbation~\cite{kubo57}. However, there is also a class of irreversible
processes induced by  thermal perturbations due to the internal inhomogeneity of a system. Among them
we have, e.g., diffusion, thermal conductivity, and viscosity. In   certain approximate schemes it is
possible to express such processes by  mechanical perturbations which artificially induce similar nonequilibrium
processes. However, the fact is that the division of perturbations into mechanical and thermal ones is
reasonable in the linear approximation only. In the higher approximations in the perturbation, 
mechanical perturbations can lead effectively to the appearance of  thermal perturbations.\\ The
NSO method permits one to formulate a workable scheme for description of the statistical mechanics of
irreversible processes which include the thermal perturbation in a unified and coherent fashion.
To perform this, it is necessary to construct  statistical ensembles representing the macroscopic conditions
determining the system. Such a formulation is quite reasonable if we consider our system for a suitable large time.
For these large times the particular properties of the initial state of the system are irrelevant and
the relevant number of variables necessary for  description of the system reduces substantially~\cite{bog}.\\
As an introduction to the NSO method, let us describe the
main ideas of this approach as follows. The basic hypothesis is that after small
time-interval $\tau$ the nonequilibrium distribution is established. Moreover, it is supposed that
it is weakly time-dependent by means of its parameter only. Then the statistical operator $\rho$ for
$t \geq \tau$ can be considered as an "integral of motion" of the quantum Liouville equation
\begin{equation}\label{eq1}
 \frac{\partial \rho }{\partial t} + \frac{1}{i \hbar}[\rho, H ] = 0
\end{equation}
Here $\frac{\partial \rho }{\partial t}$ denotes time differentiation with respect to the time variable
on which the relevant parameters $F_{m}$ depend. It is important to note once again that $\rho$ depends
on $t$ by means of $F_{m}(t)$ only. We may consider that the system is in thermal, material, and mechanical contact
with a combination of thermal baths and reservoirs  maintaining the given distribution of  parameters
$F_{m}$.  For example,  it can be the densities of energy, momentum, and particle  number for the
system which is macroscopically defined by given fields of temperature, chemical potential and
velocity. It is assumed that the chosen set of parameters is sufficient to characterize  macroscopically
the state of the system. The set of the relevant parameters are dictated by the external
conditions for the system under consideration and, therefore, the term $\frac{\partial \rho }{\partial t}$
appears as the result of the external influence upon the system. Due to this influence precisely, the
behavior of the system is nonstationary.\\ In order to describe the nonequilibrium process, it is
necessary also to choose the reduced set of relevant operators $P_{m}$, where $m$ is the index ( continuous 
or discrete). In the quantum case, all operators are considered to be in the Heisenberg representation
\begin{equation}\label{eq2}
 P_{m}(t) = \exp ( \frac{iHt}{\hbar})P_{m}\exp ( \frac{-iHt}{\hbar})
\end{equation}
where $H$ does not depend on the time. The relevant operators may be scalars or vectors. The equations
of motions for  $P_{m}$ will lead to the suitable "evolution equations"~\cite{zub74}. In the quantum
case
\begin{equation}\label{eq3}
\frac{\partial P_{m}(t)  }{\partial t} + \frac{1}{i \hbar}[P_{m}(t) , H ] = 0. 
\end{equation}
The time argument of the operator $P_{m}(t)$ denotes the Heisenberg
representation with the Hamiltonian $H$ independent of time.
Then we suppose that the state of the ensemble
is  described by a nonequilibrium statistical operator which is a functional of $P_{m}(t)$
\begin{equation}\label{eq4}
  \rho(t ) = \rho \{\ldots P_{m}(t)  \ldots \}
\end{equation}
Then $\rho(t )$ satisfies the Liouville equation (\ref{eq1}). Hence the  quasi-equilibrium ( "local-equilibrium")
Gibbs-type distribution will have the form
\begin{equation}\label{eq5}
 \rho_{q} = Q^{-1}_{q} \exp \left(  - \sum_{m}F_{m}(t)P_{m}\right)
\end{equation}
where the parameters $F_{m}(t)$ have the sense of time-dependent thermodynamic parameters,
e.g., of temperature, chemical potential, and velocity ( for the hydrodynamic stage), or the
occupation numbers of one-particle states (for the kinetic stage). The statistical
functional $Q_{q}$ is defined by demanding that the operator $\rho_{q}$ be normalized and  equal
to
\begin{equation}\label{eq6}
 Q_{q} = Tr \exp \left(  - \sum_{m}F_{m}(t)P_{m}\right)
\end{equation}
This description is still very simplified. There are various effects which can make the picture
more complicated. The quasi-equilibrium distribution is not necessarily close to the stationary
stable state. There exists another, completely independent method for choosing a suitable
quasi-equilibrium distribution~\cite{kal72,zbig,grand87,grand88,macl89}. For the state with the extremal
value of the informational entropy~\cite{grand88,macl89}
\begin{equation}\label{eq7}
  S = - Tr ( \rho \ln \rho),
\end{equation}
provided that
\begin{equation}\label{eq8}
   Tr ( \rho P_{m} ) = < P_{m}>_{q}; \quad  Tr   \rho  = 1,
\end{equation}
it is possible to construct a suitable quasi-equilibrium ensemble.
Then the corresponding quasi-equilibrium ( or local equilibrium ) distribution  has
the form
\begin{eqnarray}\label{eq9}
\rho_{q}   =   \exp \left( \Omega - \sum_{m}F_{m}(t)P_{m}\right) \equiv \exp ( S(t,0))\\
\Omega = \ln Tr  \exp \left( - \sum_{m}F_{m}(t)P_{m}\right)\nonumber
\end{eqnarray}
where $S(t,0)$ can be called  the entropy operator.
The form of the quasi-equilibrium statistical operator was constructed in so as to
ensure  that the thermodynamic equalities for the relevant parameters $F_{m}(t)$
\begin{equation}\label{eq10}
  \frac{\delta \ln Q_{q}}{\delta F_{m}(t)} = \frac{\delta \Omega}{\delta F_{m}(t)} =  - < P_{m}>_{q};
\quad \frac{\delta S}{\delta < P_{m}>_{q} }  =  F_{m}(t) 
\end{equation}
are satisfied. It is clear that the variables $ F_{m}(t)$ and $< P_{m}>_{q} $ are thermodynamically
conjugate.
Here the notation used is  $ < \ldots >_{q} =  Tr ( \rho_{q}  \ldots )$.\\
It is clear, however, that the operator $\rho_{q}$ itself  does not satisfy the
Liouville equation. The quasi-equilibrium operator should be modified in such a way that the
resulting statistical operator satisfies the Liouville equation. This is the most delicate and subtle point
of the whole method.\\  By definition a special set of operators should be constructed which depends
on the time through the parameters $F_{m}(t)$ by taking the \emph{invariant part} of the operators
$F_{m}(t)P_{m}$ occurring in the logarithm of the quasi-equilibrium distribution, i.e.,
\begin{eqnarray}\label{eq11}
 B_{m}(t) = \overline{F_{m}(t)P_{m}} = \varepsilon \int^{0}_{-\infty} e^{\varepsilon t_{1}}
F_{m}(t+ t_{1})P_{m}(t_{1})dt_{1} = \\
F_{m}(t)P_{m} - \int^{0}_{-\infty} dt_{1} e^{\varepsilon t_{1}}\left(F_{m}(t+ t_{1})\dot{P}_{m}(t_{1}) +
\dot{ F}_{m}(t+ t_{1})P_{m}(t_{1}) \right)\nonumber
\end{eqnarray}
where $(\varepsilon \rightarrow 0)$ and 
$$ \dot{P}_{m} =  \frac{1}{i \hbar}[P_{m}, H ]; \quad  \dot{ F}_{m}(t) = \frac{d F_{m}(t)}{dt}.$$
The parameter $\varepsilon > 0$ will be set
equal to zero, but only after the thermodynamic limit has been taken. Thus, the invariant part is taken
with respect to the motion with Hamiltonian $H$.
The operators $B_{m}(t)$ satisfy the Liouville equation  in the limit $(\varepsilon \rightarrow 0)$
\begin{equation}\label{eq12}
\frac{\partial B_{m}  }{\partial t} + \frac{1}{i \hbar}[ B_{m}, H ] =  
\varepsilon \int^{0}_{-\infty}  dt_{1}e^{\varepsilon t_{1}}\left(F_{m}(t+ t_{1})\dot{P}_{m}(t_{1}) +
\dot{ F}_{m}(t+ t_{1})P_{m}(t_{1}) \right)  
\end{equation}
The operation of taking the invariant part, of smoothing the oscillating terms, is used in the
formal theory of scattering~\cite{gell} to set the boundary conditions which exclude the advanced solutions
of the Schrodinger equation~\cite{watson}. It is most clearly seen when the parameters $F_{m}(t)$ are
independent of time. \\ Differentiating
$\overline{P_{m}}$ with respect to time gives
\begin{equation}\label{eq13}
\frac{\partial \overline{ P_{m}(t)}  }{\partial t}  =    \varepsilon \int^{0}_{-\infty} e^{\varepsilon t_{1}}
\dot{P}_{m}(t+t_{1})dt_{1}  
\end{equation}
The $\overline{P_{m}(t)}$ will be called the integrals ( or quasi-integrals ) of motion, although
they are conserved only in the limit $(\varepsilon \rightarrow 0)$. It is clear that for the Schrodinger equation
such a procedure excludes the advanced solutions by  choosing the initial conditions. In the
present context this procedure leads to the selection of the retarded solutions of the Liouville equation.
This philosophy has been pressed by the necessity of a consistent description of the irreversibility 
which is, according to~\cite{cov88}, " at once a profound and an elusive concept" ( c.f.,  a discussion in
Refs.~\cite{leb99,lamb}).\\ It should be noted that the same calculations can also be made with a  deeper
concept, the methods of quasi-averages~\cite{qbog,zub70,zub74}. 
Let us note once again that the quantum Liouville equation, like the classical one, is symmetric under
time-reversal transformation. However, the solution of the Liouville equation is unstable with respect 
to small perturbations violating this symmetry of the equation. Indeed, let us consider the Liouville equation
with an infinitesimally small source into the right-hand side
\begin{equation}\label{eq14}
 \frac{\partial \rho_{\varepsilon}  }{\partial t} + \frac{1}{i \hbar}[\rho_{\varepsilon}, H ] = 
 - \varepsilon ( \rho_{\varepsilon}  -  \rho_{q})
\end{equation}
or equivalently
\begin{equation}\label{eq14a}
 \frac{\partial \ln \rho_{\varepsilon}  }{\partial t} + \frac{1}{i \hbar}[ \ln \rho_{\varepsilon}, H ] = 
 - \varepsilon ( \ln \rho_{\varepsilon}  - \ln \rho_{q}),
\end{equation}
where $(\varepsilon \rightarrow 0)$ after the thermodynamic limit. This equation (\ref{eq14}) is analogous
to the corresponding equation of the quantum scattering theory~\cite{gell,watson}. The introduction
of infinitesimally small sources into  the Liouville equation is equivalent to the boundary condition
\begin{equation}\label{eq15}
  e^(\frac{iHt_{1}}{\hbar})\left( \rho(t + t_{1})  -  
  \rho_{q}(t + t_{1})  \right) e^(\frac{-iHt_{1}}{\hbar}) \rightarrow 0,
\end{equation}
where $t_{1} \rightarrow -\infty$ after the thermodynamic limiting process. It was shown~\cite{zub70,zub74}
that the operator $\rho_{\varepsilon}$ has the form
\begin{equation}\label{eq16}
\rho_{\varepsilon}(t,t) = \varepsilon \int^{t}_{-\infty} dt_{1} 
e^{\varepsilon (t_{1} - t)} \rho_{q}( t_{1},t_{1})  =
\varepsilon \int^{0}_{-\infty} dt_{1} e^{\varepsilon t_{1}}\rho_{q}(t + t_{1}, t + t_{1}) 
\end{equation}
Here the first argument of $\rho(t,t)$ is due to the indirect time-dependence via the parameters
$F_{m}(t)$ and the second one is due to the Heisenberg representation.
The required nonequilibrium statistical operator is defined as
\begin{equation}\label{eq17}
\rho_{\varepsilon}  = \rho_{\varepsilon}(t,0)   =  \overline {\rho_{q}(t,0)} = 
\varepsilon \int^{0}_{-\infty} dt_{1} 
e^{\varepsilon t_{1}} \rho_{q}(t + t_{1}, t_{1})   
\end{equation}
Hence the nonequilibrium statistical operator can then be written in the form
\begin{eqnarray}\label{eq18}
 \rho = Q^{-1} \exp \left(  - \sum_{m} B_{m} \right)   =
 Q^{-1} \exp \left(  - \sum_{m} \varepsilon \int^{0}_{-\infty}  dt_{1}e^{\varepsilon t_{1}} 
 \left( F_{m}(t+ t_{1})P_{m}(t_{1}) \right) \right)  =    
\\ Q^{-1} \exp \left(  - \sum_{m} F_{m}(t) P_{m} +
\sum_{m} \int^{0}_{-\infty}  dt_{1}e^{\varepsilon t_{1}} [ \dot{F}_{m}(t+ t_{1})P_{m}(t_{1}) + 
  F_{m}(t+ t_{1}) \dot{P}_{m}(t_{1})] \right)  \nonumber
\end{eqnarray}
Let us write down  Eq.(\ref{eq14a}) in the following form:
\begin{equation}\label{eq18a}
  \frac{d}{dt}\left( e^{\varepsilon t} \ln  \rho (t,t) \right) = \varepsilon e^{\varepsilon t}\ln  \rho_{q} (t,t),
\end{equation}
where
\begin{equation}\label{eq18b}
  \ln  \rho (t,t) = U^{\dagger}(t,0)  \ln  \rho (t,0) U(t,0); \quad U(t,0) = \exp ( \frac{iHt}{\hbar})
\end{equation}
After integration, Eq.(\ref{eq18a}),  over the interval  $(-\infty, 0)$ we get
\begin{equation}\label{18c}
  \ln  \rho (t,t) =  
  \varepsilon \int^{0}_{-\infty} dt_{1} e^{\varepsilon t_{1}} \ln \rho_{q}(t + t_{1}, t + t_{1})  
\end{equation}
Here we suppose  that $ \lim_{\varepsilon \rightarrow 0^{+}} \ln  \rho (t,t) = 0$.\\ Now we can
rewrite the nonequilibrium statistical operator in the following useful form:
\begin{equation}\label{eq18d}
    \rho (t,0) =  \exp \left(
 - \varepsilon \int^{0}_{-\infty} dt_{1} e^{\varepsilon t_{1}} \ln \rho_{q}(t + t_{1},  t_{1})  \right) = 
 \exp \overline{\left( \ln \rho_{q}(t,0)\right)} \equiv \exp \overline{\left( -S(t,0)\right)}
\end{equation}
The average value of any dynamic variable $A$ is given by
\begin{equation}\label{eq18e}
  <A> = \lim_{\varepsilon \rightarrow 0^{+}} Tr ( \rho (t,0) A )
\end{equation}
and is, in fact, the quasi-average. The normalization of the quasi-equilibrium distribution $\rho_{q}$ will
persists after taking the invariant part if the following conditions are required
\begin{equation}\label{eq18f}
  Tr ( \rho (t,0) P_{m} )= < P_{m}> = < P_{m}>_{q} ; \quad  Tr   \rho  = 1  
\end{equation}
Before closing this section, we shall mention some modification  of the "canonical"  NSO  method
which was proposed in~\cite{zbig} and which one has to take into account in a more accurate treatment
of transport processes.
%
%
\subsection{The Transport and Kinetic Equations}\label{nsok}
%
It is well known that the kinetic equations are of great interest in the theory of transport
processes. Indeed, as it was shown in the preceding section, the main quantities involved are
the following thermodynamically conjugate values:
\begin{equation}\label{eq19a}
 < P_{m}> = - \frac{\delta \Omega}{\delta F_{m}(t)};
\quad F_{m}(t)  = \frac{\delta S}{\delta < P_{m}> }   
\end{equation}
The generalized transport equations which describe the time evolution of variables $< P_{m}>$ and $F_{m}$
follow from the equation of motion for the $ P_{m}$, averaged with the nonequilibrium statistical
operator (\ref{eq18d}). It reads
\begin{equation}\label{eq19b}
 < \dot {P}_{m}> = - \sum_{n} \frac{\delta^{2} \Omega}{\delta F_{m}(t)\delta F_{n}(t)}\dot{F}_{n}(t);
\quad \dot{F}_{m}(t)  =  \sum_{n} \frac{\delta^{2} S}{\delta < P_{m}> \delta < P_{n}>} <\dot {P}_{n}>  
\end{equation}
The entropy production has the form
\begin{equation}\label{eq19c}
\dot {S}(t) = < \dot {S}(t,0)> = - \sum_{m}< \dot {P}_{m}> F_{m}(t) = -
\sum_{n,m}\frac{\delta^{2} \Omega }{\delta F_{m}(t)\delta F_{n}(t)}\dot{F}_{n}(t)F_{m}(t)
\end{equation}
These equations are the mutually conjugate and  with   Eq.(\ref{eq19a}) form a complete system of equations
for the calculation of values  $< P_{m}>$ and $F_{m}$.\\
Let us illustrate the NSO method by considering the derivation of kinetic equations for a system of 
weakly interacting particles~\cite{pok68}. In this case the Hamiltonian can be written in the form
\begin{equation}\label{eq19}
 H = H_{0} + V,
\end{equation}
where $H_{0}$ is the Hamiltonian of noninteracting particles ( or quasiparticles ) and $V$ is the
operator describing the weak interaction among them. Let us choose the set of operators $ P_{m} = P_{k}$
whose average values correspond to the particle distribution functions, e.g.,  $a^{\dagger}_{k}a_{k}$ or
$a^{\dagger}_{k}a_{k+q}$. Here $a^{\dagger}_{k}$ and $a_{k}$ are the creation and annihilation second
quantized operators ( Bose or Fermi type). These operators obey the following quantum equation of
motion:
\begin{equation}\label{eq20}
 \dot{P}_{k} = \frac{1}{i \hbar}[P_{k} , H ]
\end{equation}
It is reasonable to assume that the following relation is fulfilled
\begin{equation}\label{eq21}
 [P_{k} , H_{0} ] = \sum_{l} c_{kl}P_{l},
\end{equation}
where $c_{kl}$ are some coefficients ( c-numbers).\\
According to  Eq.(\ref{eq18}), the nonequilibrium statistical operator has the form
\begin{eqnarray}\label{eq22}
 \rho =  Q^{-1} \exp \left(  - \sum_{k} F_{k}(t) P_{k} +
\sum_{k} \int^{0}_{-\infty}  dt_{1}e^{\varepsilon t_{1}} [ \dot{F}_{k}(t+ t_{1})P_{k}(t_{1}) + 
  F_{k}(t+ t_{1}) \dot{P}_{k}(t_{1})] \right)   
\end{eqnarray}
After elimination of the time-derivatives with the help of the equation $<P_{k}> = <P_{k}>_{q}$
it can be shown~\cite{pok68} that the integral term in the exponent, Eq.(\ref{eq22}), will be
proportional to the interaction $V$. The averaging of Eq.(\ref{eq20}) with NSO (\ref{eq22}) gives the
generalized kinetic equations for $<P_{k}>$
\begin{equation}\label{eq24}
 \frac{d <P_{k}>  }{d t} = \frac{1}{i \hbar}<[  P_{k} , H ]> =
 \frac{1}{i \hbar} \sum_{l} c_{kl}<P_{l}> +  \frac{1}{i \hbar}<[  P_{k} , V ]>
\end{equation}
Hence the calculation of the r.h.s. of (\ref{eq24})
leads to the explicit expressions for the "collision integral" ( collision terms). 
Since the interaction is small, it is possible to rewrite Eq.(\ref{eq24}) in the following form:
\begin{equation}\label{eq25}
 \frac{d <P_{k}>  }{d t} = L^{0}_{k} + L^{1}_{k} + L^{21}_{k} + L^{22}_{k},
 \end{equation}
where
\begin{equation}\label{eq26}
 L^{0}_{k} =  \frac{1}{i \hbar} \sum_{l} c_{kl}<P_{l}>_{q}
\end{equation}
\begin{equation}\label{eq27}
 L^{1}_{k} =  \frac{1}{i \hbar} <[  P_{k} , V ]>_{q}
\end{equation}
\begin{equation}\label{eq28}
 L^{21}_{k} =  \frac{1}{ \hbar^{2}}  \int^{0}_{-\infty} dt_{1} e^{\varepsilon t_{1}} < [ V(t_{1}), [ P_{k} , V ]]>_{q}
\end{equation}
\begin{equation}\label{eq29}
 L^{22}_{k} =  \frac{1}{ \hbar^{2}}  \int^{0}_{-\infty} dt_{1} 
 e^{\varepsilon t_{1}} < [ V(t_{1}),  i \hbar \sum_{l} P_{l}   
 \frac{\partial   L^{1}_{k}(\ldots <P_{l}> \ldots)  }{\partial  <P_{l}>} ]>_{q}
\end{equation}
The higher order terms  proportional to the $V^{3}$, $V^{4}$, etc., can be derived straightforwardly.
%
\subsection{Kinetic Equations for Magnons and Phonons}\label{nsom}
%
The dynamic  behavior of charge~\cite{reg},  magnetic~\cite{spark}, and lattice~\cite{pei} systems is of interest for the study 
of transport processes in solids. Partial emphasis has been placed on the derivation of the kinetic equations
describing the  hot electron transport in semiconductors~\cite{kal70,xing}, and the relaxation  of 
magnons~\cite{sh61,phi66} and phonons~\cite{pei,car} due to the inelastic scattering of  
quasiparticles. \\
We discuss briefly in this section the processes occurring after the switching off the external
magnetic field in  a  ferromagnetic crystal. Our main interest is in ferromagnetic insulators,
where the dominant interaction is the Heisenberg exchange coupling $ J \vec{S}_{i}\vec{S}_{j}$.
It is well known that a strong microwave magnetic field applied parallel
to the dc field can give rise to parametric excitation of spin 
waves~\cite{sh61,esh61,sh62}. In this technique the wave number of the
potentially unstable spin waves can be changed by varying the dc magnetic field. One thus obtains information
about the variation of the spin-wave relaxation time with the wave number~\cite{sh63,esh63,sh66,sh67,sh69}.
Because of its relative  simplicity the "parallel pumping" technique has proved very useful in determining
rather fundamental properties of ferromagnetic materials. The subharmonic generation of spin waves at high power levels
is an efficient research tool for probing magnon-magnon and magnon-phonon interactions. Useful information
about the spin-wave relaxation rate can be deduced from the kinetic equations to study magnon-magnon and
magnon-phonon interactions.\\ Here the spin-wave relaxation processes arising from the dipolar interaction
will be considered as an  example. The Hamiltonian has the form
\begin{equation}\label{eq30}
 H = - \frac{1}{2}\sum_{l  \neq l'} J(R_{ll'} )\vec{S_{l}}\vec{S_{l'}} + 2\mu_{0}h_{0}\sum_{l}S^{z}_{l} +
2\mu_{0}^{2} \sum_{l  \neq l'} \frac{1}{R^{5}_{ll'}} \left( R^{2}_{ll'} \vec{S_{l}} \vec{S_{l'}} - 
3 (\vec{R_{ll'}} \vec{S_{l'}}) (\vec{R_{ll'}} \vec{S_{l'}}) \right)
\end{equation}
This Hamiltonian contains Zeeman energy, exchange energy, and dipolar energy. To treat this Hamiltonian,
it should be expressed in terms of the amplitudes of the normal modes or spin waves~\cite{tyab67}. The amplitudes
of the normal modes are  quantum-mechanically interpreted as creation and annihilation operators ( usually
bosons). 
We get~\cite{tyab67}
\begin{equation}\label{eq31}
 S^{+}_{l} =  S^{x}_{l} +  iS^{y}_{l} =  
 \sqrt{2S} b^{\dagger}_{l} \sqrt{1 - \frac{b^{\dagger}_{l} b_{l}}{2S}}; \quad 
S^{-}_{l} =  S^{x}_{l} -  iS^{y}_{l} =  
 \sqrt{2S}  \sqrt{1 - \frac{b^{\dagger}_{l} b_{l}}{2S} b_{l}}; \quad  S^{z}_{l} = - S + b^{\dagger}_{l} b_{l}
\end{equation}
We adopt the notation
$$b_{i} =
N^{-1/2}\sum_{\vec k}  b_{k} \exp (i{\vec k} {\vec R_{i}}),
\quad b^{\dagger}_{i} = N^{-1/2}\sum_{\vec k}
b^{\dagger}_{k} \exp (-i{\vec k} {\vec R_{i}})$$
The transformed Hamiltonian  contains a term that is quadratic in the spin-wave amplitudes $H^{(2)}$ and also terms that are of higher order,
 $H^{(3)}$, $H^{(4)}$, etc. 
\begin{equation}\label{eq32}
  H = H^{(2)} + H^{(3)} + H^{(4)} + \ldots
\end{equation}
The eigenstates of the quadratic part of the Hamiltonian $H^{(2)}$ can be characterized by the occupation
numbers $c^{\dagger}_{k}c_{k}$, i.e., the quadratic part can be diagonalized to the form~\cite{tyab67}
\begin{equation}\label{eq33}
   H^{(2)} = \sum_{k} \epsilon(k)c^{\dagger}_{k}c_{k}; \quad \epsilon(k) = \hbar \omega(k)
\end{equation}
Here the operators $c^{\dagger}_{k}$  and $c_{k}$ are the second-quantized operators of creation and 
annihilation of magnons.
All higher order terms in the Hamiltonian lead to transitions between the eigenstates. 
In terms of the magnon operators these terms are given by
\begin{equation}\label{eq34}
   H^{(3)} = \sum_{kpp'}\Phi(k,p,p')c^{\dagger}_{k}c^{\dagger}_{p} c_{p'}\Delta (\vec{k}+ \vec{p} - 
   \vec{p'}) + H. C.
\end{equation}
\begin{equation}\label{eq35}
   H^{(4)} = \sum_{kpp'r}\Phi(k,p;p',r)c^{\dagger}_{k}c^{\dagger}_{p} c_{p'} c_{r}\Delta (\vec{k}+ \vec{p} - 
\vec{p'} - \vec{r})     + H. C.
\end{equation}
Usually,  only
that term in the Hamiltonian which is of the third order in the amplitudes of the normal modes
is considered explicitly, because only this term leads to the relaxation rates proportional to the
temperature in the high-temperature limit.\\ Let us apply now the formalism of generalized kinetic equations,
as described above. We suppose that the set of averages $<P_{k}> = <c^{\dagger}_{k}c_{k}> = <n_{k}>$ characterize the 
nonequilibrium state of the system. The quasi-equilibrium statistical operator   has the form
\begin{equation}\label{eq36}
 \rho_{q} =  Q^{-1}_{q} \exp \left(  - \sum_{k} F_{k}(t)n_{k}\right); 
 \quad  Q^{-1}_{q} = Tr\exp \left(  - \sum_{k} F_{k}(t)n_{k}\right)
\end{equation}
The kinetic equation (\ref{eq25}) can then be expressed by
\begin{equation}\label{eq37}
 \frac{d <n_{k}>  }{d t} = L^{0}_{k} + L^{1}_{k} + L^{21}_{k} + L^{22}_{k}
 \end{equation}
For both the contributions $H^{(3)}$  and  $H^{(4)}$ the following equality holds:
\begin{equation}\label{eq38}
 L^{0}_{k} = L^{1}_{k} = L^{22}_{k}  = 0
 \end{equation}
Let us first consider the contribution of the term $H^{(3)}$. We can then write
\begin{eqnarray}\label{eq39}
L^{21}_{k} =   - \frac{8\pi}{ \hbar} \sum_{\vec{p},\vec{p'} } \{|\Phi(k,p,p')|^{2} 
\delta \left( \omega(k) + \omega(p) - \omega(p')\right)\Delta (\vec{k}+ \vec{p} - \vec{p'})  \\ \nonumber
[( <n_{k}> + 1)( <n_{p}> + 1)<n_{p'}>  - <n_{k}><n_{p}>( <n_{p'}> + 1)]  \\ \nonumber
- \frac{1}{2}|\Phi(k,p,p')|^{2} 
\delta \left( \omega(k) - \omega(p) - \omega(p')\right)\Delta (\vec{k}- \vec{p} - \vec{p'})  \\ \nonumber
[( <n_{k}> + 1)<n_{p}><n_{p'}>   - <n_{k}>(<n_{p}> + 1 )( <n_{p'}> + 1)]
\end{eqnarray}
We can make the same calculation to obtain $L^{21}_{k}$ for the  magnon-magnon scattering term $H^{(4)}$
\begin{eqnarray}\label{eq40}
L^{21}_{k} =   - \frac{16\pi}{ \hbar} \sum_{\vec{p},\vec{p'},\vec{r} } \{|\Phi(k,p,p',r)|^{2} 
\delta \left( \omega(k) + \omega(p) - \omega(p') -  \omega(r) \right)\Delta (\vec{k}+ \vec{p} - \vec{p'} -
\vec{r}  )  \\ \nonumber
[( <n_{k}> + 1)( <n_{p}> + 1)<n_{p'}><n_{r}>  - <n_{k}><n_{p}>( <n_{p'}> + 1)( <n_{r}> + 1)]  
\end{eqnarray}
Here the notation was used
$$<n_{k}> = N(\hbar \omega(k)) = [\exp (\beta \hbar \omega(k)) - 1]^{-1}$$
The quantities $\Phi(k,p,p')$ and $\Phi(k,p,p',r)$ are the combination of the matrix elements which
describe the various transitions between spin eigenstates~\cite{sh61}.  Equation (\ref{eq39}) 
 corresponds precisely to the rate equation which describes the change of the average occupation 
 number $<n_{k}>$ of the mode $k$ derived in~\cite{sh61}. The discussion of the two relevant relaxation
 rates $\tau^{-1}_{k}$ ( due to the confluence and splitting ) is given there. The types of kinetic 
 equations,  Eqs.(\ref{eq39}), (\ref{eq40}),
 involved in our derivation and the conclusions arrived at show very clearly that the NSO method is a workable
 and useful approach for  derivation of the  kinetic equations for  concrete physical problems. As far as  
 the kinetic equations for magnons is concerned, its convenience can become even more evident if one needs to
 take into account higher order magnon processes ( four, five, etc.). The higher order processes may give rise to  additional
 and unusual behavior ( i.e.,  a general heating of the spin-wave system causing the saturation, additional
 smaller peaks and kinks in the measured curves, etc.)\\
It is evident that a similar derivation can be given for the  kinetic equation for   phonons. The
theory of thermal conductivity~\cite{car,lep} has been  extensively developed  beginning with the
kinetic theory of Peierls~\cite{pei74,pei,pei29}. The theory of lattice thermal conductivity invented by Peierls~\cite{pei,pei29}
is based on the assumption that the perturbing mechanisms to the harmonic case 
( anharmonicity, imperfections) are small in magnitude.
The Peierls collision term for the three-phonon processes $H^{(3)}$  looks like
\begin{eqnarray}\label{eq41}
L^{21}_{k}  \sim    \frac{a\pi}{ \hbar} \sum_{\vec{p},\vec{p'} } \{|\Phi(k,p,p')|^{2} 
\delta \left( \omega(k) + \omega(p) - \omega(p')\right)  \\ \nonumber
[( <n_{k}> + 1)( <n_{p}> + 1)<n_{p'}>  - <n_{k}><n_{p}>( <n_{p'}> + 1)]  \\ \nonumber
+ \frac{1}{2}|\Phi(k,p,p')|^{2} 
\delta \left( \omega(k) - \omega(p) - \omega(p')\right) 
[( <n_{k}> + 1)<n_{p}><n_{p'}>  ]
\end{eqnarray}
Note  that our calculations show that three- and four-phonon processes behave quite differently.
One expects that the stronger the anharmonicity the larger the thermal resistance. To catch this trend, some
sophisticated formalisms~\cite{mac65} have been developed which utilize a modified version of 
the Peierls-Boltzmann equation.  A useful approach to improving the initial Peierls theory corresponds to the derivation
of a generalized Peierls-Boltzmann equation, where the phonons in the collision term are treated not as free phonons but as
 quasiparticles with a finite width and damping which are determined self-consistently.  Crystal
lattices at low temperatures represent an interacting system of quasiparticles  in which we observe two relaxation
mechanisms of widely different time scales, i.e., the system either at short or long times after its initial
perturbation from equilibrium. For the long-time behavior of the system it is possible to formulate the
problem in terms of the correlation functions of quantities relaxing slowly, such as densities of conserved
variables in the system. The corresponding transport equations are similar in structure to the phonon
Boltzmann equation with a modification of the collision term.  A detailed study of the  transport equations
for phonon systems   is not within the scope of this paper and deserves a separate consideration.

\section{ System in Thermal Bath: Generalized Kinetic Equations}\label{nsob}
%
We now proceed to derive  generalized kinetic equations for  the system weakly coupled to a thermal bath.
Examples of such systems can be an atomic ( or molecular) system interacting with the electromagnetic
field it generates as with a thermal bath, a system of electrons or exitons interacting with the phonon field,
etc.
Our aim is to investigate  relaxation processes in two weakly interacting subsystems, one of which is in the
nonequilibrium state and the other is considered as a thermal bath. The concept of thermal bath or heat reservoir,
i.e.,  a system that has effectively an infinite number of degrees of freedom, was not formulated precisely. A 
standard definition of the thermal bath is a heat reservoir defining a temperature of the system environment.
From a mathematical point of view~\cite{bog1}, a heat bath is something that gives a stochastic influence on the system
under consideration. In this sense, the generalized master equation~\cite{vanh,leaf} is a tool for extracting the dynamics of
a subsystem of a larger system  by the use of a special projection techniques~\cite{zwan64}. The problem of
a small system weakly interacting with a heat reservoir has various aspects. For example, a useful model
of the lattice thermal conduction is a problem of a stationary energy current through a crystalline
lattice in contact with external heat reservoirs~\cite{ches,leb59,pasta}. Basic to the derivation of a
transport equation for a small system weakly interacting with a heat bath is a proper introduction of  model assumptions.\\
We are interested here in the problem of
derivation of the kinetic equations for a certain set of  average values 
( occupation numbers, spins, etc.)  which characterize the nonequilibrium state of the system.\\
Let us consider the relaxation of a small subsystem weakly interacting with a thermal bath. The Hamiltonian of the
total system is taken in the following form:
\begin{equation}\label{eq42}
 H = H_{1} + H_{2} + V,
\end{equation}
where
\begin{equation}\label{eq43}
 H_{1} = \sum_{\alpha} E_{\alpha}a^{\dagger}_{\alpha}a_{\alpha}; 
 \quad V = \sum_{\alpha,\beta}\Phi_{\alpha \beta}a^{\dagger}_{\alpha}a_{ \beta}, \quad \Phi_{\alpha \beta} = 
 \Phi^{\dagger}_{\alpha \beta}
\end{equation}
Here $H_{1}$ is the Hamiltonian of the small subsystem, and $a^{\dagger}_{\alpha}$ and $a_{\alpha}$ are the creation and annihilation second
quantized operators of quasiparticles in the small subsystem with energies $ E_{\alpha}$, $V$ is the
operator of the interaction between the small subsystem and the thermal bath, and $H_{2}$ is the
Hamiltonian of the thermal bath  which we do not write explicitly. The quantities $\Phi_{\alpha \beta}$ are
the operators acting on the thermal bath variables.\\ We are interested in the kinetic stage of the
nonequilibrium process in the system weakly coupled to a thermal bath. Therefore, we assume that the state of this
system is determined completely by the set of averages
 $<P_{\alpha \beta}> = <a^{\dagger}_{\alpha}a_{ \beta}>$ and the state of the thermal bath by $<H_{2}>$,
 where $ < \ldots >$ denotes the statistical average with the nonequilibrium statistical operator,
 which will be defined below.\\
In order to pursue our discussion, we will use
the whole development in section \ref{nso}. We take the quasi-equilibrium statistical operator $\rho_{q}$ in
the form
\begin{eqnarray}\label{eq44}
\rho_{q} (t) =  \exp (- S(t,0)), \quad  S(t,0) = \Omega(t) + 
\sum_{\alpha \beta }P_{\alpha \beta }F_{\alpha\beta }(t) + \beta H_{2}\\ 
\Omega = \ln Tr \exp (- \sum_{\alpha \beta }P_{\alpha \beta }F_{\alpha\beta }(t) - \beta H_{2} ) \nonumber
\end{eqnarray}
Here $F_{\alpha\beta }(t)$ are the thermodynamic parameters conjugated with $P_{\alpha \beta }$, and $\beta$ is
the reciprocal temperature of the thermal bath. All the operators are considered in the Heisenberg
representation. The nonequilibrium statistical operator has the form 
\begin{eqnarray}\label{eq44}
\rho (t) =  \exp (- \overline{S(t,0)}),\\
\overline{S(t,0)} = \varepsilon \int^{0}_{-\infty} dt_{1} e^{\varepsilon t_{1}} 
\left( \Omega(t + t_{1}) +  \sum_{\alpha \beta }P_{\alpha \beta }F_{\alpha\beta }(t) + \beta H_{2}  \right)
\nonumber
\end{eqnarray}
The parameters $F_{\alpha\beta }(t)$ are determined from the condition 
$<P_{\alpha \beta }> = <P_{\alpha \beta }>_{q}$.\\ In the derivation of the kinetic equations we  use
the perturbation theory in a "weakness of interaction" and  assume that the equality 
$ <\Phi_{\alpha \beta}>_{q} = 0$ holds, while  other terms can be added to the renormalized energy of
the subsystem. The nonequilibrium statistical operator can be rewritten as
\begin{eqnarray}\label{eq45}
\rho (t) = Q^{-1} \exp (- L(t)),\\
L(t)   = \varepsilon \int^{0}_{-\infty} dt_{1} e^{\varepsilon t_{1}} 
\left(  \sum_{\alpha \beta }P_{\alpha \beta }F_{\alpha\beta }(t + t_{1}) + \beta H_{2}(t_{1})  \right)
\nonumber
\end{eqnarray}
Integrating in Eq.(\ref{eq45}) by parts, we obtain
\begin{eqnarray}\label{eq46}
L(t) = \sum_{\alpha \beta } P_{\alpha \beta } F_{\alpha \beta }(t) + \beta H_{2}  \\
- \int^{0}_{-\infty} dt_{1} e^{\varepsilon t_{1}}
\left( \sum_{\alpha \beta } \dot{P}_{\alpha \beta }(t_{1}) F_{\alpha \beta}(t + t_{1}) +  
\sum_{\alpha \beta} P_{\alpha \beta }(t_{1}) \frac{\partial F_{\alpha \beta }(t + t_{1})}{\partial t_{1}}
+ \beta \dot{H _{2}}(t_{1})\right)  \nonumber
\end{eqnarray}
For further considerations it is convenient to rewrite $\rho_{q}$ as
\begin{equation}\label{eq47}
 \rho_{q}  = \rho _{1}\rho_{2} = Q^{-1}_{q} \exp (- L_{0}(t)),
\end{equation}
where
\begin{eqnarray}\label{eq48}
\rho _{1} = Q^{-1}_{1}\exp \left(- \sum_{\alpha \beta } P_{\alpha \beta } F_{\alpha \beta }(t)\right); \quad
Q_{1} = Tr \exp \left( - \sum_{\alpha \beta } P_{\alpha \beta } F_{\alpha \beta }(t) \right)\\
\rho _{2} = Q^{-1}_{2} e^{- \beta H_{2} }; \quad Q_{2} = Tr \exp (- \beta H_{2})\\
Q_{q} = Q_{1}Q_{2}; \quad L_{0} = \sum_{\alpha \beta } P_{\alpha \beta } F_{\alpha \beta }(t) + \beta H_{2}
\end{eqnarray}
We now turn to the derivation of the kinetic equations. The starting point is the kinetic equations
in the following implicit form:
\begin{equation}\label{eq49}
 \frac{d <P_{\alpha \beta }> }{d t} =  \frac{1}{i \hbar}<[  P_{\alpha \beta  } , H ]> =
 \frac{1}{i \hbar}(E_{\beta} - E_{\alpha})<P_{\alpha \beta  }> +  \frac{1}{i \hbar}<[  P_{\alpha \beta} , V ]>
\end{equation}
We restrict ourselves to the second-order in powers of $V$ in calculating the r.h.s. of (\ref{eq49}). To
this end, we must obtain $\rho (t)$ in the first-order in $V$. We get
\begin{eqnarray}\label{eq50}
\frac{\partial F_{\alpha \beta }(t + t_{1})}{\partial t_{1}} = \frac{i}{ \hbar}(E_{\beta} - E_{\alpha})
F_{\alpha \beta }(t + t_{1}) -  \sum_{ \mu \nu }  
\frac{\partial F_{\alpha \beta }(t + t_{1})}{\partial <P_{\mu \nu}> }
 \frac{1}{i \hbar}<[  P_{ \mu \nu}(t_{1}) , V(t_{1}) ]> = \nonumber \\
\frac{i}{ \hbar}(E_{\beta} - E_{\alpha})
F_{\alpha \beta }(t + t_{1}) -  \sum_{ \mu \nu \gamma}  \frac{\partial F_{\alpha \beta }(t + t_{1})}{\partial <P_{\mu \nu}> }
\left( < \Phi_{ \nu \gamma}P_{\mu \gamma } >  -    <\Phi_{\gamma \mu} P_{\gamma \nu  } > \right) 
\end{eqnarray}
Restricting ourselves to the linear terms in Eq.(\ref{eq50}), we obtain
\begin{eqnarray}\label{eq51}
\frac{\partial F_{\alpha \beta }(t + t_{1})}{\partial t_{1}} \simeq \frac{i}{\hbar}(E_{\beta} - E_{\alpha})
F_{\alpha \beta }(t + t_{1}) -  \frac{1}{i \hbar} \sum_{ \mu \nu }  
\frac{\partial F_{\alpha \beta }(t + t_{1})}{\partial <P_{\mu \nu}> }
\left( < \Phi_{ \nu \gamma}>_{q} <P_{\mu \gamma } >  -    <\Phi_{\gamma \mu}>_{q} < P_{\gamma \nu  } > \right) \nonumber \\
= \frac{i}{ \hbar}(E_{\beta} - E_{\alpha})F_{\alpha \beta }(t + t_{1})
\end{eqnarray}
The quantities, $\dot{P}_{\alpha \beta }(t_{1}) $ and $\dot{H_{2}(t_{1})}$ in the first-order in
interaction have the form
\begin{eqnarray}\label{eq52}
\dot{P}_{\alpha \beta }(t_{1}) = \frac{1}{i\hbar}(E_{\beta} - E_{\alpha})P_{\alpha \beta }(t_{1}) +
 \frac{1}{i \hbar} [  P_{\alpha \beta  } ,  V(t_{1}) ]  \nonumber \\
\dot{H}_{2}(t_{1}) =  \frac{1}{i \hbar} [ H_{2}(t_{1}),  V(t_{1})]
\end{eqnarray}
Here and below all the operators are taken in the interaction representation. Using 
Eqs.(\ref{eq51}) and (\ref{eq52}) we find
\begin{eqnarray}\label{eq53}
L(t) =  L_{0}  
- \int^{0}_{-\infty} dt_{1} e^{\varepsilon t_{1}}
\left[   \sum_{\alpha \beta } P_{\alpha \beta }(t_{1}) F_{\alpha \beta}(t + t_{1}) 
+  \beta H _{2}(t_{1}),  V(t_{1}) \right]
\end{eqnarray}
It can be verified that the expression $ \sum_{\alpha \beta } P_{\alpha \beta }(t_{1}) F_{\alpha \beta}(t + t_{1}) 
+  \beta H _{2}$ is independent of $t_{1}$ in the zero-order in interaction and consequently is equal 
to $L_{0}$. Then for $\rho (t)$ in the linear approximation in interaction $V$ we have
\begin{eqnarray}\label{eq54}
\rho (t) =  \rho_{q}  -  \frac{i}{\hbar} \rho_{q} \int^{0}_{-\infty} dt_{1} e^{\varepsilon t_{1}} 
\int^{1}_{0} d\lambda e^{\lambda L_{0}} [L_{0},  V(t_{1})] e^{- \lambda L_{0}}
\end{eqnarray}
By integrating in Eq.(\ref{eq54}) over  $\lambda$ and using the relation
\begin{eqnarray}\label{eq55}
e^{\lambda L_{0}} [L_{0},  V(t_{1})] e^{- \lambda L_{0}} = \frac{d}{d\lambda}
e^{\lambda L_{0}}  V(t_{1})  e^{- \lambda L_{0}}
\end{eqnarray}
we get
\begin{equation}\label{eq56}
\rho (t) =  \rho_{q}  -  \frac{i}{\hbar} \int^{0}_{-\infty} dt_{1} e^{\varepsilon t_{1}} [V(t_{1}),  \rho_{q}] 
\end{equation}
Finally, with the aid of Eq.(\ref{eq56}) we obtain the kinetic equations for $<P_{\alpha \beta }>$ in
the form
\begin{equation}\label{eq57}
 \frac{d <P_{\alpha \beta }> }{d t} =  \frac{1}{i \hbar}(E_{\beta} - E_{\alpha})<P_{\alpha \beta  }> -
 \frac{1}{\hbar^{2}} \int^{0}_{-\infty} dt_{1} e^{\varepsilon t_{1}} 
 < \left[[P_{\alpha \beta}, V], V(t_{1}) \right]>_{q}
\end{equation}
The last term of the right-hand side of Eq.(\ref{eq57}) can be called the generalized "collision integral".
Thus,  we can see that the collision term for the system  weakly coupled to the thermal bath has a convenient
form of the double commutator as for the generalized kinetic equations (\ref{eq28}) for the system with
small interaction. It should be emphasized  that the assumption about the model form of the Hamiltonian
(\ref{eq42}) is nonessential for the above derivation. We can start again with the Hamiltonian (\ref{eq42}) 
in which we shall not specify the explicit form of $H _{1}$ and $V$. We assume that the state of the
nonequilibrium system is characterized completely by some set of  average values $<P_{k}>$ and the
state of the thermal bath by $<H _{2}>$. We confine ourselves to such systems for 
which $[H _{1}, P_{k}] =  \sum_{l} c_{kl}P_{l}$. Then we assume that $<V>_{q} \simeq 0$, where $<\ldots>_{q}$
denotes the statistical average with the quasi-equilibrium statistical operator of the form
\begin{eqnarray}\label{eq58}
\rho_{q}  
 = Q^{-1}_{q} \exp \left(  - \sum_{k} P_{k}F_{k}(t) 
-  \beta H _{2} \right) 
\end{eqnarray}
and $F_{k}(t)$ are the parameters conjugated with $<P_{k}>$. Following the method used above in the derivation 
of  equation (\ref{eq57}), we can obtain the generalized kinetic equations for $<P_{k}>$ with an accuracy up to
terms which are quadratic in interaction
\begin{equation}\label{eq59}
 \frac{d <P_{k}> }{d t} =  \frac{i}{ \hbar} \sum_{l} c_{kl}<P_{l}> -
 \frac{1}{\hbar^{2}} \int^{0}_{-\infty} dt_{1} e^{\varepsilon t_{1}} 
 < \left[[P_{k}, V], V(t_{1}) \right]>_{q}
\end{equation}
Hence (\ref{eq57}) is fulfilled for the general form of the Hamiltonian of a small system weakly coupled to a
thermal bath.
%
\section{ System in Thermal Bath: Balance and Master Equations}\label{nsobm}
In  section  \ref{nsob} we have obtained the kinetic equations for $<P_{\alpha \beta }>$ in
the general form. Our next task is to write down   equations (\ref{eq57}) in an explicit form. To do this,
we note that the perturbation operator can be represented as $V(t_{1}) = 
\sum_{\alpha,\beta} \phi_{\alpha \beta}(t_{1})a^{\dagger}_{\alpha}a_{ \beta}$, 
where 
\begin{equation}\label{eq60}
 \phi_{\alpha \beta}(t_{1}) =   U_{2}(t_{1} )  \Phi_{\alpha \beta}  
 U^{\dagger}_{2}(t_{1} )\exp (  \frac{i}{ \hbar}(E_{\alpha} - E_{\beta})t_{1}); 
 \quad U_{2}(t_{1}) = \exp ( \frac{iH_{2}t_{1}}{\hbar})
\end{equation}
Now  we calculate the double commutator in the right-hand side of Eq.(\ref{eq57})
\begin{eqnarray}\label{eq61}
 < \left[[P_{\alpha \beta}, V], V(t_{1}) \right]>_{q} = 
 \\   \sum_{\mu \nu } 
  \{ < \Phi_{ \beta \mu} \phi_{\mu \nu}(t_{1})>_{q}<P_{\alpha \nu }> +
< \phi_{\nu \mu}(t_{1}) \Phi_{\mu \alpha}>_{q}<P_{\nu \beta }>   -    
(< \Phi_{\mu \alpha} \phi_{\beta \nu}(t_{1})>_{q}  + < \phi_{\mu \alpha}(t_{1}) 
\Phi_{\beta \nu }>_{q}) <P_{\mu \nu }>  \} \nonumber
\end{eqnarray}
where we restricted ourselves to the linear terms in the mean density of quasiparticles. 
Let us now remind that the correlation functions $ < A B(t)>$  and  $ < A(t) B>$ can be expressed 
via their spectral intensities. Indeed,
an effective way of viewing quasiparticles,  quite general and
consistent, is via the Green functions scheme of many-body
theory\cite{tyab67,kuzem02}.  It is known~\cite{tyab67,zub74}
that the correlation functions and Green functions are completely determined by the spectral weight
function ( or spectral intensity) $J(\omega)$. 
\begin{eqnarray}
\label{eq62} F_{AB} (t-t') = < A (t) B(t') > =  \frac {1}{2\pi}
\int^{ + \infty}_{ - \infty} d\omega
\exp [i\omega (t - t')] J_{AB} (\omega)  \\
F_{BA} (t'-t) = < B(t') A(t) > =  \frac {1}{2\pi} \int^{ +
\infty}_{ - \infty} d\omega \exp [i\omega (t'-t)] J_{BA} (\omega)
\label{eq63}
\end{eqnarray}
Here  the Fourier transforms  $J_{AB}(\omega)$  and
$J_{BA}(\omega)$ are of the form
\begin{eqnarray}
\label{eq64}
J_{BA} (\omega) = \\
Q^{-1} 2 \pi \sum_{m,n} \exp(-\beta E_{n}) (\psi^{\dagger}_{n} B
\psi_{m})
(\psi^{\dagger}_{m} A \psi_{n}) \delta ( \frac{E_{n} - E_{m}}{\hbar} - \omega )\nonumber   \\
J_{AB}( - \omega) = \exp( \beta \hbar\omega)J_{BA} (\omega) \label{eq65}
\end{eqnarray}
Expressions (\ref{eq64}) and (\ref{eq65}) are  spectral
representations of the corresponding time correlation functions.
The quantities $J_{AB}$ and $J_{BA}$ are  spectral
densities or spectral weight functions. \\
It is convenient to define
\begin{eqnarray}
\label{eq66} F_{BA} (0) = < B(t) A(t) > =  \frac {1}{2\pi}
\int^{ + \infty}_{ - \infty} d\omega
J (\omega)  \\
F_{AB} (0) = < A (t) B(t) > =  \frac {1}{2\pi} \int^{ + \infty}_{
- \infty} d\omega \exp (\beta \hbar\omega ) J (\omega) \label{eq67}
\end{eqnarray}
The correlation functions $ < \Phi_{ \beta \mu} \phi_{\mu \nu}(t_{1})>_{q}$ 
and $< \phi_{\nu \mu}(t_{1}) \Phi_{\mu \alpha}>_{q}$ are connected with their spectral intensities in the
following way:
\begin{eqnarray}
\label{eq72}
< \Phi_{\mu \nu} \phi_{\gamma \delta}(t)>_{q} = \frac {1}{2\pi}\int^{ + \infty}_{ - \infty} d\omega 
J_{\gamma \delta,\mu \nu} (\omega) \exp [-i( \omega - \frac{E_{\gamma} - E_{\delta}}{ \hbar})t] \\
< \phi_{ \mu \nu}(t) \Phi_{\gamma \delta}>_{q}  = \frac {1}{2\pi}\int^{ + \infty}_{ - \infty} d\omega 
J_{\gamma \delta,\mu \nu} (\omega) \exp [i( \omega + \frac{E_{\mu} - E_{\nu}}{ \hbar})t] 
\label{eq73}
\end{eqnarray}
Substituting Eqs.(\ref{eq72}) and (\ref{eq73}) into Eqs.(\ref{eq57}) and (\ref{eq61}) and taking into 
account the notation
\begin{eqnarray}
\label{eq74}
 \frac{1}{i\hbar}\sum_{\mu} 
 \int^{ 0}_{ - \infty} dt_{1}  e^{\varepsilon t_{1}}   < \Phi_{ \beta \mu} \phi_{\mu \nu}(t_{1})>_{q} =
\frac{1}{2\pi}  \sum_{\mu}  \int^{ + \infty}_{ - \infty} d\omega
 \frac{J_{\mu \nu, \beta \mu} (\omega)}{\hbar\omega - E_{\gamma} - E_{\delta} + i\varepsilon} = K_{\beta\nu}\\
 \frac{1}{i\hbar} \int^{ 0}_{ - \infty} dt_{1}  e^{\varepsilon t_{1}} 
(< \Phi_{ \mu \alpha} \phi_{\beta \nu}(t_{1})>_{q} + 
< \phi_{\mu \alpha}(t_{1}) \Phi_{\beta \nu}>_{q})  = \nonumber \\
\frac{1}{2\pi}   \int^{ + \infty}_{ - \infty} d\omega J_{\beta \nu, \mu \alpha} (\omega)
 \left( \frac{1}{\hbar\omega - E_{\beta} + E_{\nu} + i\varepsilon}  
 - \frac{1}{\hbar\omega - E_{\alpha} - E_{\mu} - i\varepsilon}\right)  = K_{\alpha \beta,\mu \nu} 
\label{eq75}
\end{eqnarray}
one can rewrite the kinetic equations for $< P_{\alpha \beta }>$ as
\begin{equation}\label{eq76}
 \frac{d <P_{\alpha \beta }> }{d t} =  \frac{1}{i \hbar}(E_{\beta} - E_{\alpha})<P_{\alpha \beta  }> -
\sum_{\nu} \left( K_{\beta\nu}<P_{\alpha \nu }>  + K^{\dag}_{\alpha \nu} <P_{\nu \beta}> \right) +
K_{\alpha \beta,\mu \nu} <P_{\mu \nu }>
\end{equation}
If one confines himself to the diagonal averages $< P_{\alpha \alpha}>$ only, the last equation 
may be transformed to give
\begin{equation}\label{eq77}
 \frac{d <P_{\alpha \alpha }> }{d t} =  \sum_{\nu} K_{\alpha \alpha,\nu \nu} <P_{\nu \nu }>
 - \left( K_{\alpha \alpha}  + K^{\dag}_{\alpha \alpha}  \right)<P_{\alpha \alpha }>
\end{equation}
\begin{eqnarray}\label{eq78}
K_{\alpha \alpha,\nu \nu} =  \frac{1}{\hbar^{2}} J_{\alpha \nu, \nu \alpha} ( \frac{E_{\alpha} - E_{\beta}}{\hbar} ) = 
W_{\beta \rightarrow \alpha}\\
 K_{\alpha \alpha}  + K^{\dag}_{\alpha \alpha}  = 
 \frac{1}{\hbar^{2}}\sum_{\nu}  J_{\nu \alpha, \alpha \nu} ( \frac{E_{\beta} - E_{\alpha}}{\hbar} ) = 
W_{\alpha \rightarrow \beta} 
\end{eqnarray}
Here $W_{\beta \rightarrow \alpha}$ and $W_{\alpha \rightarrow \beta}$ are the transition probabilities
expressed in the spectral intensity terms. Using the properties of the spectral intensities~\cite{tyab67},
it is possible to verify that the transition probabilities satisfy the relation of the detailed
balance
\begin{equation}\label{eq79}
\frac{W_{\beta \rightarrow \alpha}}{W_{\alpha \rightarrow \beta}} = 
\frac{\exp (-\beta E_{\alpha})}{\exp (-\beta E_{\beta})}
\end{equation}
Finally, we have
\begin{equation}\label{eq80}
 \frac{d <P_{\alpha \alpha }> }{d t} =  \sum_{\nu} W_{\nu \rightarrow \alpha}  <P_{\nu \nu }>
 - \sum_{\nu} W_{\alpha \rightarrow \nu}  <P_{\alpha \alpha}>
 \end{equation}
This equation has the usual form of the Pauli master equation. \\ According to Ref.~\cite{lu91}, "the master equation
is an ordinary differential equation, describing the {\em reduced evolution} of the system, obtained from the full
Heisenberg evolution by taking the partial expectation with respect to the vacuum state of the reservoirs degrees
of freedom".
The rigorous mathematical derivation of the generalized master equation~\cite{lu91,vanh,leaf,zwan64,swen,pet,ful} is
rather a complicated mathematical problem. 
%
%
\section{A Dynamical System in a Thermal Bath}
\label{tb} 
%
The problem about the appearance of a stochastic process in a dynamical system  which is submitted
to the influence of a "large" system was considered by Bogoliubov~\cite{bog1,bog2}. For a classical system this
question was studied on the basis of the Liouville equation for the probability distribution in the phase
space  and for quantum mechanical systems on the basis of an analogous equation for the von Neumann  
statistical operator. In the mentioned papers a mathematical method was elaborated which permitted  
obtaining, in the first approximation, the Fokker-Planck equations. Since then  a lot of papers were devoted
to studying this problem from  various points of view ( e.g. Refs.~\cite{rub,kor,g1,g3,kas,op,g4}). Lebowitz and 
Rubin~\cite{rub} studied the motion of a Brownian particle in a fluid ( as well as the  motion 
of a Brownian particle in a crystal ) from a dynamical point of view. They derived a formal structure of the
collision term similar to the structure of the usual linear transport equation. Kassner~\cite{kas} used a new type
of projection operator and derived homogeneous equations of motion for the reduced density operator of a system
coupled to a bath. It was shown that in order to consistently describe damping within quantum mechanics,  one must couple
the open system of interest to a heat reservoir. 
The problem of the inclusion of dissipative forces in quantum mechanics
is of  great interest. There are various approaches to this complicated problem~\cite{lu91,kor,kub89,petr,sen}.
Tanimura and Kubo~\cite{kub89} considered a
test system coupled to a bath system with linear interactions and derived a set of hierarchical equations for the
evolution of their reduced density operator. Breuer and Petruccione~\cite{petr}  developed a formulation of quantum statistical
ensembles in terms of probability distributions on a projective Hilbert state. They derived a Liouville master
equation for the reduced probability distribution of an open quantum system. It was shown that the time-dependent wave 
function of an open quantum system represented a well-defined stochastic process which is generated by the
nonlinear Schrodinger equation
\begin{equation}\label{eq80a}
  \frac{\partial \psi}{\partial t} = -i G(\psi)
\end{equation}
with the nonlinear and non-Hermitian operator $G(\psi)$. The inclusion of dissipative forces in quantum mechanics
through the use of non-Hermitian Hamiltonians is of  great interest in the theory of  interaction between heavy ions.
It is clear that if the Hamiltonian has a non-Hermitian part $H_{A}$ the Heisenberg equation of motion will be modified
by  additional terms. However, care must be taken in defining the probability density operator when the 
Hamiltonian is non-Hermitian. Also, the state described by the wave function $\psi$ is  not  then  an energy eigenstate
because of the energy dissipation. As it was formulated by Accardi and Lu~\cite{lu91}, "the quantum Langevin equation is
a quantum stochastic differential equation driven by some quantum noise ( creation, annihilation, number noises)." 
The necessity of considering such  processes arises in the description of various quantum phenomena ( e.g.,
radiation damping, etc.), since quantum systems experience dissipation and fluctuations through interaction with
a reservoir~\cite{lax1,lax2}.
The concept of "quantum noise" was proposed by Senitzky~\cite{sen} to derive a quantum dissipation mechanism.
Originally, the time evolution of quantum systems with the dissipation and fluctuations was described by adding
a dissipative term to the quantum equation of motion. However, as was noted by Senitzky~\cite{sen}, this procedure
leads to the nonunitary time evolution. He proposed to derive the quantum dissipation mechanism by 
introducing   {\em quantum noise}, i.e., a  quantum field interacting with the dynamical 
system (in his case  an oscillator). For an appropriately chosen form of the interaction, energy will flow away
from the oscillator to the quantum noise field (thermal bath or reservoir).\\
In this section, we consider the behavior of
a small dynamic  system interacting with a thermal bath, i.e., with a system that has effectively an infinite
number of degrees of freedom, in the approach of the nonequilibrium statistical operator, on the basis of the equations 
derived in section \ref{nsob}. The equations derived below can help in the understanding of the origin of
irreversible behavior in quantum phenomena.\\ 
We  assume that the dynamic  system ( system of particles) is far from
equilibrium with the thermal bath and cannot, in general, be characterized by a temperature. As a result of
the interaction with the thermal bath, such a system   acquires some statistical characteristics but   remains
essentially a mechanical system. Our aim is to obtain an equation of evolution ( equations of motion ) for the
relevant variables which are characteristic of the system under consideration. The basic idea is to
eliminate effectively the thermal bath variables (c.f. Ref.~\cite{swe,lax1,lax2} ).  The influence of the thermal bath
is manifested then as an effect of friction of the particle in a medium. The presence of friction leads to
dissipation and, thus, to irreversible processes. In this respect,  our philosophy coincides precisely with the Lax 
statement~\cite{lax1}  "that the reservoir can be {\em completely eliminated} provided that the frequency shifts and dissipation
induced by the reservoir are incorporated into the mean equations of motion, and provided that a suitable operator noise source with 
the correct moments are added".\\
Let us consider the behavior of a small subsystem with Hamiltonian
$H_{1}$ interacting with a thermal bath with Hamiltonian $H_{2}$. The total Hamiltonian has the form 
(\ref{eq42}). As operators $P_{m }$ determining the nonequilibrium state of the small subsystem, we take
$a^{\dagger}_{\alpha}, a_{\alpha}$, and $n_{\alpha} = a^{\dagger}_{\alpha}a_{\alpha}$. Note  that the choice
of only the operators $n_{\alpha}$ and $H_{2}$ would lead to kinetic equations (\ref{eq59})  for the system in the thermal bath
derived above.\\
The quasi-equilibrium statistical operator (\ref{eq5}) is determined from the extremum of the
information entropy (\ref{eq7}) subject to the additional conditions that the quantities
\begin{equation}\label{eq81}
Tr (\rho a_{\alpha}) = <a_{\alpha}>, \quad Tr (\rho a^{\dagger}_{\alpha}) = <a^{\dagger}_{\alpha}>, \quad
Tr (\rho n_{\alpha}) = <n_{\alpha}>
\end{equation}  
remain constant during the variation and the normalization $Tr (\rho ) = 1$ is preserved. The operator
$\rho_{q}$ has the form
\begin{eqnarray}\label{eq82}
\rho_{q}   =   \exp \left( \Omega - \sum_{\alpha} ( f_{\alpha}(t)a_{\alpha} + f^{\dagger}_{\alpha}(t)a^{\dagger}_{\alpha} + 
F_{\alpha}(t)n_{\alpha} ) - \beta H_{2} \right) \equiv \exp ( S(t,0))\\
\Omega = \ln Tr  \exp \left( - \sum_{\alpha}( f_{\alpha}(t)a_{\alpha} + f^{\dagger}_{\alpha}(t)a^{\dagger}_{\alpha} + 
F_{\alpha}(t)n_{\alpha} ) - \beta H_{2}  \right)\nonumber
\end{eqnarray}
Here, $f_{\alpha}, f^{\dagger}_{\alpha}$ and $F_{\alpha}$ are Lagrangian multipliers determined by the conditions
(\ref{eq81}). They are the parameters conjugate to $<a_{\alpha}>_{q}, <a^{\dagger}_{\alpha}>_{q}$  
and $<n_{\alpha}>_{q}$:
\begin{equation}\label{eq83}
<a_{\alpha}>_{q} =  - \frac{\delta \Omega}{\delta f_{\alpha}(t)},  \quad <n_{\alpha}>_{q} =
- \frac{\delta \Omega}{\delta F_{\alpha}(t)},
\quad \frac{\delta S}{\delta < a_{\alpha}>_{q} }  =  f_{\alpha}(t),  \quad
\frac{\delta S}{\delta < n_{\alpha}>_{q} } = F_{\alpha}(t) 
\end{equation}  
In what follows, it is convenient to write the quasi-equilibrium statistical operator (\ref{eq82}) in the
form
\begin{equation}\label{eq84}
\rho_{q} = \rho_{1} \rho_{2},    
\end{equation}
where
\begin{eqnarray}\label{eq85}
\rho_{1}   =   \exp \left( \Omega_{1} - \sum_{\alpha} ( f_{\alpha}(t)a_{\alpha} + f^{\dagger}_{\alpha}(t)a^{\dagger}_{\alpha} + 
F_{\alpha}(t)n_{\alpha} ) \right) \\
\Omega_{1} = \ln Tr  \exp \left( - \sum_{\alpha}( f_{\alpha}(t)a_{\alpha} + f^{\dagger}_{\alpha}(t)a^{\dagger}_{\alpha} + 
F_{\alpha}(t)n_{\alpha} ) \right)\nonumber \\
\rho_{2}   =   \exp \left( \Omega_{2}  - \beta H_{2} \right), \quad
\Omega_{2} = \ln Tr  \exp \left( - \beta H_{2}  \right)
\label{eq86} 
\end{eqnarray}
The nonequilibrium statistical operator $\rho$ will have the form (\ref{eq18d}). Note, that the following conditions
are satisfied:
\begin{equation}\label{eq87}
<a_{\alpha}>_{q} = <a_{\alpha}>, \quad <a^{\dagger}_{\alpha}>_{q} = <a^{\dagger}_{\alpha}>, \quad
<n_{\alpha}>_{q} = <n_{\alpha}>
\end{equation}  
We shall take, as our starting point, the equations of motion for the operators averaged with the
nonequilibrium statistical operator (\ref{eq18d}) 
\begin{eqnarray}\label{eq88}
i\hbar \frac{d <a_{\alpha}>}{dt} = < [a_{\alpha}, H_{1} ]> + < [a_{\alpha}, V ]>,\\
i\hbar \frac{d <n_{\alpha}>}{dt} = < [n_{\alpha}, H_{1} ]> + < [n_{\alpha}, V ]>
\label{eq89} 
\end{eqnarray}
The equation for $<a^{\dagger}_{\alpha}>$ can be obtained by taking the conjugate of  (\ref{eq88}). Restricting
ourselves to the second order in the interaction V, we obtain, by analogy with (\ref{eq59}), the following
equations:
\begin{eqnarray}\label{eq90}
i\hbar \frac{d <a_{\alpha}>}{dt} = E_{\alpha}< a_{\alpha}> + \frac{1}{i\hbar }
\int^{0}_{-\infty} dt_{1} e^{\varepsilon t_{1}} 
 < \left[[a_{\alpha}, V], V(t_{1}) \right]>_{q}\\
i\hbar \frac{d <n_{\alpha}>}{dt} = 
\frac{1}{i\hbar }
\int^{0}_{-\infty} dt_{1} e^{\varepsilon t_{1}} 
 < \left[[n_{\alpha}, V], V(t_{1}) \right]>_{q}
\label{eq91} 
\end{eqnarray}
Here $V(t_{1})$ denotes the interaction representation of the operator $V$. Expanding the double commutator
in Eq.(\ref{eq90}), we obtain
\begin{equation}\label{eq92}
i\hbar \frac{d <a_{\alpha}>}{dt} = E_{\alpha}< a_{\alpha}> + \frac{1}{i\hbar }
\int^{0}_{-\infty} dt_{1} e^{\varepsilon t_{1}} \left( \sum_{\beta \mu \nu} 
<\Phi_{\alpha \beta}\phi_{\mu \nu}(t_{1})>_{q}< a_{\beta}a^{\dagger}_{\mu} a_{\nu}>_{q}   -
<\phi_{\mu \nu}(t_{1})\Phi_{\alpha \beta}>_{q}<a^{\dagger}_{\mu} a_{\nu} a_{\beta}>_{q} \right),
\end{equation}
where $\phi_{\mu \nu}(t_{1}) = \Phi_{\mu \nu}(t_{1}) \exp ( \frac{i}{\hbar }(E_{\mu} - E_{\nu})t_{1} )$. We transform
Eq.(\ref{eq92}) to  
\begin{eqnarray}\label{eq93}
i\hbar \frac{d <a_{\alpha}>}{dt} = E_{\alpha}< a_{\alpha}> + \frac{1}{i\hbar }\sum_{\beta \mu }
\int^{0}_{-\infty} dt_{1} e^{\varepsilon t_{1}} <\Phi_{\alpha \mu}\phi_{\mu \beta}(t_{1})>_{q}< a_{\beta}> + \\
\frac{1}{i\hbar }\sum_{\beta \mu \nu}
\int^{0}_{-\infty} dt_{1} e^{\varepsilon t_{1}} <[\Phi_{\alpha \nu}, \phi_{\mu \nu}(t_{1})]>_{q}
<a^{\dagger}_{\mu} a_{\nu} a_{\beta}>_{q} \nonumber
\end{eqnarray}
We  assume that the terms of higher order than linear can be dropped in Eq.(\ref{eq93})  ( below, 
we shall formulate the conditions when this is possible). Then we get
\begin{equation}\label{eq94}
  i\hbar \frac{d <a_{\alpha}>}{dt} = E_{\alpha}< a_{\alpha}> + \frac{1}{i\hbar }\sum_{\beta \mu }
\int^{0}_{-\infty} dt_{1} e^{\varepsilon t_{1}} <\Phi_{\alpha \mu}\phi_{\mu \beta}(t_{1})>_{q}< a_{\beta}>
\end{equation}
The form of the linear equation (\ref{eq94}) is the same for Bose and Fermi statistics.\\ Using the 
spectral representations,  Eq.(\ref{eq72}) and Eq.(\ref{eq73}), it is possible  to rewrite Eq.(\ref{eq94}) 
by analogy with Eq.(\ref{eq76}) as
\begin{equation}\label{eq95}
  i\hbar \frac{d <a_{\alpha}>}{dt} = E_{\alpha}< a_{\alpha}> + \sum_{\beta }
K_{\alpha \beta} < a_{\beta}>
\end{equation}
where $K_{\alpha \beta} $ is defined in (\ref{eq74}). Thus, we have obtained the equation of
motion for the average $<a_{\alpha}>$. 
It is clear that this equation describes approximately the evolution of the state of the dynamic  system interacting with
the thermal bath. The last term in the right-hand side of this equation leads to the shift of 
energy $E_{\alpha}$ and to the damping due to the interaction  with the thermal bath ( or medium ). In a certain sense,
it is possible to say that  Eq.(\ref{eq95}) is an analog or the generalization of the 
Schrodinger equation.\\ Let us now show how, in the case of Bose statistics, we can take into account the
nonlinear terms which lead to a coupled system of equations for $<a_{\alpha}>$ and $<n_{\alpha}>$. Let us consider
the quantity $<a^{\dagger}_{\mu} a_{\nu} a_{\beta}>_{q}$. After the canonical transformation
$$ a_{\alpha}  = b_{\alpha} + <a_{\alpha}>, \quad  a^{\dagger}_{\alpha}  = b^{\dagger}_{\alpha} + <a^{\dagger}_{\alpha}>
$$
the operator $\rho_{1} $ in Eq.(\ref{eq85}) can be written in the form
\begin{equation}\label{eq97}
 \rho_{1}   =  Q^{-1}_{1} \exp \left( \Omega_{1} - \sum_{\alpha} ( 
F_{\alpha}(t) b^{\dagger}_{\alpha}b_{\alpha} ) \right), 
\quad  <a_{\alpha}> = - \frac{f^{\dagger}_{\alpha}}{F_{\alpha}}
\end{equation}
Note that $Q_{1}$ in (\ref{eq97}) is not, in general, equal to $Q_{1}$ in (\ref{eq85}). Using 
the Wick-De Dominicis theorem~\cite{tyab67} for the operators $b^{\dagger}_{\alpha}, b_{\alpha}$ and
returning to the original operators $a^{\dagger}_{\alpha}, a_{\alpha}$, we obtain
\begin{equation}\label{eq98}
<a^{\dagger}_{\mu} a_{\nu} a_{\beta}>_{q}  \simeq  (<n_{\mu}> -  
|<a_{\mu}>|^{2} )<a_{\nu}>\delta_{\mu,\beta} + (<n_{\mu}> -  
|<a_{\mu}>|^{2} )<a_{\beta}>\delta_{\mu,\beta} 
\end{equation}
Using (\ref{eq98}), we can rewrite Eq.(\ref{eq85}) in the form
\begin{eqnarray}\label{eq99}
i\hbar \frac{d <a_{\alpha}>}{dt} = E_{\alpha}< a_{\alpha}> + \frac{1}{i\hbar }\sum_{\beta \mu }
\int^{0}_{-\infty} dt_{1} e^{\varepsilon t_{1}} <\Phi_{\alpha \mu}\phi_{\mu \beta}(t_{1})>_{q}< a_{\beta}> + \\
\frac{1}{i\hbar }\sum_{ \mu \beta}
\int^{0}_{-\infty} dt_{1} e^{\varepsilon t_{1}} \left( <[\Phi_{\alpha \mu}, \phi_{\mu \beta}(t_{1})]>_{q} +
<[\Phi_{\alpha \beta}, \phi_{\mu \mu}(t_{1})]>_{q}
\right) (<n_{\mu}> +  |<a_{\mu}>|^{2} )<a_{\beta}>  \nonumber
\end{eqnarray}
Now consider Eq.(\ref{eq91}). Expand the double commutator and, in the same way as the threefold terms
were neglected in the derivation of  Eq.(\ref{eq94}), ignore the fourfold terms in (\ref{eq91}). We obtain
then
\begin{eqnarray}\label{eq100}
\frac{d <n_{\alpha}>}{dt} = \sum_{\beta}W_{\beta \rightarrow \alpha}(<n_{\beta}> +  |<a_{\beta}>|^{2} ) -
\sum_{\beta}W_{\alpha \rightarrow \beta}(<n_{\alpha}> +  |<a_{\alpha}>|^{2} ) +\\
\frac{1}{i\hbar }\sum_{\beta} K_{\alpha \beta} <a^{\dagger}_{\alpha}>< a_{\beta}> + 
\frac{1}{i\hbar }\sum_{\beta} K^{\dagger}_{\alpha \beta} <a_{\alpha}>< a^{\dagger}_{\beta}> +
\sum_{\mu \nu}K_{\alpha \alpha,\mu \nu} <a^{\dagger}_{\mu}>< a_{\nu}> \nonumber
\end{eqnarray}
Thus, in the general case Eqs.(\ref{eq90}) and (\ref{eq91}) form a coupled system of nonlinear equations
of Schrodinger and kinetic types. The nonlinear equation (\ref{eq92}) of Schrodinger type is an auxiliary
equation and, in conjunction with the equation of kinetic type (\ref{eq100}), determines the parameters of the
nonequilibrium statistical operator since in the case of Bose statistics
\begin{eqnarray}\label{eq101}
<a_{\alpha}> = - \frac{f^{\dagger}_{\alpha}(t)}{F_{\alpha}(t)}, 
\quad <n_{\alpha}> = (e^{ F_{\alpha}(t)} - 1)^{-1} + \frac{|f_{\alpha}|^{2}}{F_{\alpha}^{2}(t)}
\end{eqnarray}
Therefore, the linear Schrodinger equation is a fairly good approximation if
$$(<n_{\alpha}> +  |<a_{\alpha}>|^{2} ) = (e^{ F_{\alpha}(t)} - 1)^{-1} \ll 1 $$  The last condition
corresponds essentially to $<b^{\dagger}_{\alpha}b_{\alpha}>  \ll  1$.\\ In the case of Fermi statistics the
situation is more complicated~\cite{ber}. There is well-known isomorphism between bilinear products of
fermion operators and the Pauli spin matrices~\cite{kae}.  In quantum field theory the sources linear in the Fermi operators are 
introduced by means of classical spinor fields that anticommute with one another and with the original field.
The Fermion number processes in the time evolution of a certain quantum Hamiltonian model were 
investigated in Ref.~\cite{lu93}. It was shown that the time evolution tended to the solution of a quantum stochastic 
differential equation driven by the Fermion number processes.
We shall not consider here this complicated case.\\ In order to interpret the physical meaning of the derived
equations, an example will be given here. Let us consider briefly a system of electrons in a lattice described by the
Hamiltonian
\begin{equation}\label{eq102}
H = H_{1} + H_{2} + V = \sum_{k\sigma}\epsilon(k)a^{\dagger}_{k\sigma} a_{k\sigma } + 
\sum_{q} \hbar \omega_{q} b^{\dagger}_{q} b_{q} + \frac{1}{\sqrt{v}}\sum_{k_{1},k_{2}\sigma}
A(\vec{k_{1}} - \vec{k_{2}})a^{\dagger}_{k_{1} \sigma} a_{k_{2} \sigma }(b_{\vec{k_{1}} - \vec{k_{2}} } + 
b^{\dagger}_{\vec{k_{2}} - \vec{k_{1}} } ),
\end{equation}
where $\hbar \omega_{q} $ is the phonon energy, $a^{\dagger}_{k\sigma},  a_{k\sigma }$ 
and $b^{\dagger}_{q}, b_{q}$ are the operators of creation and annihilation of electrons and phonons, respectively;
$\epsilon(k)$ is the energy of electrons and $ A(\vec{q})$ determines the electron-phonon coupling. Equation
(\ref{eq95}) for $<a_{k\sigma }>$  can be represented in the form
\begin{equation}\label{eq103}
  i\hbar \frac{d <a_{k\sigma }>}{dt} = (\epsilon(k) + \Delta E(k))< a_{k\sigma }> -  
  \frac{i\hbar}{2}\Gamma(k)< a_{k\sigma }>,
\end{equation}
where
\begin{eqnarray}\label{eq104}
\Delta E(k) = P \sum_{k_{1}} | A(\vec{k} - \vec{k_{1}})|^{2} 
\left( \frac{ <N_{k - k_{1}}> + 1}
{\epsilon(k) - \epsilon(k_{1}) - \hbar \omega_{\vec{k} - \vec{k_{1}}}} + 
\frac{<N_{k  - k_{1}}> }
{\epsilon(k) - \epsilon(k_{1}) + \hbar \omega_{\vec{k} - \vec{k_{1}}}} 
\right)\\
\Gamma(k) = \frac{2\pi}{\hbar} \sum_{k_{1}}| A(\vec{k} - \vec{k_{1}})|^{2} 
\left( (<N_{k - k_{1}}> + 1) \delta (\epsilon(k) - \epsilon(k_{1}) - \hbar \omega_{\vec{k} - \vec{k_{1}}})
 + <N_{k - k_{1}}>  \delta (\epsilon(k) - \epsilon(k_{1}) + \hbar \omega_{\vec{k} - \vec{k_{1}}})\right)
\label{eq105}
\end{eqnarray}
are the energy shift of an electron and the electron damping, respectively. 
Here $ <N_{q}> = (e^{\beta \hbar \omega_{q} } - 1)^{-1}$, the distribution functions of the phonons.  
Expressions (\ref{eq104}) and (\ref{eq105}) are the same as those obtained by the Green functions
method~\cite{zub60} if one sets $<a^{\dagger}_{k\sigma} a_{k\sigma }> \ll 1$ in the latter.
%
%
\section{ Schrodinger-Type Equation with Damping for a Dynamical System in a Thermal Bath}
\label{seq} 
%
In the previous section we obtained an equation for   mean values of the amplitudes in the
form (\ref{eq95}). It is of interest to analyze and   track more closely the analogy with the
Schrodinger equation in the coordinate form. To do this, by convention we define the "wave function"
\begin{equation}\label{eq106}
\psi ( \vec{r}) = \sum_{\alpha} \chi_{\alpha}( \vec{r})<a_{\alpha}>,
\end{equation}
where $\{ \chi_{\alpha}( \vec{r}) \}$ is a complete orthonormalized system of single-particle functions of the
operator
$\left( - \frac{\hbar^{2}}{2m}\nabla^{2}  + v(\vec{r}) \right)$,
where $v(\vec{r})$ is the potential energy, and
\begin{equation}\label{eq107}
\left(- \frac{\hbar^{2}}{2m}\nabla^{2}  + v(\vec{r}) \right)\chi_{\alpha}( \vec{r}) = 
E_{\alpha} \chi_{\alpha}( \vec{r})
\end{equation}
Thus, in a certain sense, the quantity $\psi ( \vec{r})$ may plays the role of the wave function of a  
particle in the medium. Now, using (\ref{eq106}), we transform Eq.(\ref{eq95}) to  
\begin{equation}\label{eq108}
 i\hbar \frac{\partial \psi ( \vec{r}) }{\partial t} = 
 \left(- \frac{\hbar^{2}}{2m} \nabla^{2}  + v(\vec{r}) \right) \psi( \vec{r})  +
 \int K (\vec{r},\vec{r'})\psi ( \vec{r'})d \vec{r'}
\end{equation}
The kernel $K (\vec{r},\vec{r'})$ of the integral equation (\ref{eq108}) has the form
\begin{equation}\label{eq109}
  K (\vec{r},\vec{r'}) = \sum_{\alpha \beta} K_{\alpha \beta}\chi_{\alpha}( \vec{r})\chi^{\dag}_{\beta}( \vec{r'}) = 
\frac{1}{i\hbar} \sum_{\alpha, \beta, \mu } \int^{0}_{-\infty} dt_{1} e^{\varepsilon t_{1}}
<\Phi_{\alpha \mu}\phi_{\mu \beta}(t_{1})>_{q}\chi_{\alpha}( \vec{r})\chi^{\dag}_{\beta}( \vec{r'})
\end{equation}
Equation (\ref{eq108}) can be called a Schrodinger-type equation with damping for a dynamical system in
a thermal bath. It is interesting to note that similar Schrodinger equations with a nonlocal interaction are used in the
scattering theory~\cite{lax,mot} to describe interaction with many scattering centers. \\To demonstrate the
capabilities of   equation (\ref{eq108}), it is convenient to introduce the operator of translation
$ \exp (i\vec{q} \vec{p} / \hbar)$, where $\vec{q} = \vec{r'} - \vec{r}; \vec{p} = -i\hbar \nabla_{r}$.
Then Eq.(\ref{eq108}) can be rewritten in the form
\begin{equation}\label{eq110}
 i\hbar \frac{\partial \psi ( \vec{r}) }{\partial t} = 
 \left(- \frac{\hbar^{2}}{2m} \nabla^{2}  + v(\vec{r}) \right) \psi( \vec{r})  +
\sum_{p} D(\vec{r},\vec{p})\psi ( \vec{r}) 
\end{equation}
where
\begin{equation}\label{eq111}
D(\vec{r},\vec{p}) = \int d^{3}q K (\vec{r},\vec{r} + \vec{q})e^{\frac{i\vec{q}\vec{p}}{\hbar} }
\end{equation}
It is reasonable to assume that the wave function $\psi ( \vec{r})$ varies little over the correlation 
length characteristic of the kernel $K (\vec{r},\vec{r'})$. Then, 
expanding $ \exp (i\vec{q} \vec{p} / \hbar)$ in a series, we obtain the following equation in the
zeroth order:
\begin{equation}\label{eq112}
 i\hbar \frac{\partial \psi ( \vec{r}) }{\partial t} = 
 \left(- \frac{\hbar^{2}}{2m} \nabla^{2}  + v(\vec{r} + \text{Re}  U( \vec{r})) \right) \psi( \vec{r})  +
 i \text{Im}  U( \vec{r}) \psi( \vec{r})
\end{equation}
where
\begin{equation}\label{eq113}
U( \vec{r}) = \text{Re}  U( \vec{r}) +  i \text{Im}  U( \vec{r}) =
 \int d^{3}q K (\vec{r},\vec{r} + \vec{q}) 
\end{equation}
Expression (\ref{eq112}) has the form of a Schrodinger equation with a complex potential. 
Equations of this form are well known in the scattering theory~\cite{mot}  in which one introduces an 
interaction describing absorption  $(\text{Im}  U( \vec{r})  < 0)$. Further, expanding 
$ \exp (i\vec{q} \vec{p} / \hbar)$ in a series up to the second order inclusively, we can represent
Eq.(\ref{eq108}) in the following form~\cite{mot}:
\begin{equation}\label{eq114}
 i\hbar \frac{\partial \psi ( \vec{r}) }{\partial t} = 
\{  \left( - \frac{\hbar^{2}}{2m} \nabla^{2}  + v(\vec{r} ) \right)  +
 U( \vec{r}) - \frac{1}{i\hbar}\int d^{3}q K (\vec{r},\vec{r} + \vec{q})(\vec{q}\vec{p} )
 +  \frac{1}{2}\int d^{3}q K (\vec{r},\vec{r} + \vec{q})  \sum^{3}_{m,n=1}q^{m} q^{n} \nabla_{m} \nabla_{n}               
 \} \psi( \vec{r})
\end{equation}
To interpret this equation, let us introduce the function
\begin{equation}\label{eq115}
 \vec{A}(\vec{r} ) = \frac{mc}{i\hbar e}\int d^{3}q   \text{Re} K (\vec{r},\vec{r} + \vec{q})\vec{q}
\end{equation}
where $m$  and $e$ are the mass and charge of the electron and $c$ is the velocity of light. 
Then $ \vec{A}(\vec{r} )$ can be considered, in a certain sense, as an analog of the complex vector
potential of an electromagnetic field. It is clear  that the motion of a particle (dynamic  subsystem)
through the medium imitates, to some extent, the motion of a charged particle in the electromagnetic field.
To make this analogy even more close, let us introduce the following quantity:
\begin{equation}\label{eq116}
\left( \frac{1}{M (\vec{r} )} \right)_{ij}
  =  \frac{1}{m}\delta_{ij} - 
 \frac{mc}{i\hbar e}\int d^{3}q K (\vec{r},\vec{r} + \vec{q}) q^{i} q^{j}
\end{equation}
It follows from (\ref{eq116}) that this quantity can be interpreted as a tensor of the 
reciprocal effective masses~\cite{bloc,mott}. The notion of the "mass tensor" was 
introduced in~\cite{bloc} to describe the motion of an electron in an external field F 
\begin{equation}\label{eq117}
\frac{dv_{i}}{dt} = \frac{e}{\hbar^{2} } \sum_{j} \frac{\partial^{2} E}{\partial q^{i} \partial q^{j}} F_{j}
\quad i,j =1,2,3 \quad \text{or} \quad x, y, z
\end{equation}
or in vector notation
\begin{equation}\label{eq118}
\frac{d \vec{v}}{dt} = \frac{e}{\hbar^{2} } \text{grad}_{q}(\vec{F} \text{grad}_{q}E)
\end{equation}
Thus,  a field $\vec{F}$ may change the velocity $\vec{v}$ in directions other than that of $\vec{F}$. The quantity
$\hbar^{2} (\frac{\partial^{2} E}{\partial q^{i} \partial q^{j}} )^{-1}  $ has been called the "mass tensor".
Now  we can rewrite Eq.(\ref{eq114}) in the form
\begin{eqnarray}\label{eq119}
 i\hbar \frac{\partial \psi ( \vec{r}) }{\partial t} = 
  \left( - \frac{\hbar^{2}}{2}  \sum^{3}_{i,j=1} \left( \frac{1}{M (\vec{r} )} \right)_{ij} 
\nabla_{i}\nabla_{j}  + v(\vec{r})  +  U( \vec{r}) +   \frac{ie\hbar}{mc}\vec{A}(\vec{r} )\vec{\nabla} + 
i T ( \vec{r}) \right)\psi ( \vec{r})\\
T ( \vec{r}) = \frac{1}{2}\int d^{3}q K (\vec{r},\vec{r} + \vec{q})  \sum^{3}_{m,n=1}q^{m} q^{n} \nabla_{m} \nabla_{n}
\end{eqnarray}
Note  that in an isotropic medium the tensor $ \left( \frac{1}{M (\vec{r} )} \right)_{ij}$ is diagonal
and $\vec{A}(\vec{r} ) = 0$.
The introduction of $\psi ( \vec{r})$ does not mean that the state of the small dynamical subsystem is
pure. It remains mixed since it is described by the statistical operator (\ref{eq18d}), the evolution of the
parameters  $f_{\alpha}, f^{\dagger}_{\alpha}$, and $F_{\alpha}$  of the latter being governed by a 
coupled system of equations of Schrodinger and kinetic types. It is interesting to mention that the derivation of a
Schrodinger-type equation with non-Hermitian  Hamiltonian which describes the dynamic  and statistical aspects of 
the motion was declared by Korringa~\cite{kor}. However, his Eq.(29)
\begin{equation}\label{eq120}
 i  \frac{\partial W' }{\partial t} =  \left ( H'(t) + h'(t)) + \frac{i}{2\theta} \frac{dh'}{dt} + \ldots \right)  W'(t)
\end{equation}
where $ W'(t)$ is the statistical matrix for the primed system,   can  hardly  be considered as a Schrodinger-type 
equation. This special form of the equation  for the time-dependent statistical matrix can be considered 
as a modified Bloch equation.\\
Hence we were able to apply the NSO approach given above to dynamics. 
We have shown in this section  that for some class of dynamic 
systems it was possible,  with the NSO approach,  to go from a Hamiltonian description of dynamics to a description in terms
of processes which incorporates the dissipativity. However, a careful examination is required in order to see under what conditions
the Schrodinger-type equation with damping can really be used.
%
%
%
%
%
%
%
\section{ Concluding Remarks} 
%
In this paper, we have discussed the general  statistical mechanics approach to the description of the transport processes.
We have applied the method of the nonequilibrium statistical operator to study the generalized kinetic and
evolution equations. We  analyzed and derived in a closed form the  kinetic
equations and applied them to some typical problems.\\
In writing the paper we have essentially confined ourselves to a discussion of those features of
the theory which deal with general structural properties rather than with specific physical
applications. The method offers several advantages over the standard technique of the calculation of
transport coefficients.
The derived generalized kinetic equations for a system weakly coupled to
a thermal bath are analogous to those derived in~\cite{pok68} for the system of  weakly interacting
particles. Moreover, the capability of the generalized kinetic equations was demonstrated and further discussed
by considering a few representative examples, i.e.,  the kinetic equations for magnons and phonons,  and
the energy shift and damping of particle (electron) due to the friction with media (phonons). There are many other applications
of the formalism developed in this article, for example, longitudinal nuclear spin relaxation and spin
diffusion. However, we have not considered other contributions here. These questions deserve a separate 
consideration.\\
An example of a small system being initially far from equilibrium has been considered.
We have reformulated the theory of the time evolution of a small dynamic  system weakly coupled to a thermal
bath and shown that a Schrodinger-type equation emerges from this theory as a particular case. Clearly then, the nonequilibrium
statistical operator approach is a convenient and workable tool for the derivation of relaxation equations and  
 formulae for evolution and kinetic equations.\\ In our above 
treatment we have avoided a number of important questions  such as the rigorous proof of the existence and uniqueness of the
quasi-equilibrium state, the validity of the time-smoothing procedure, etc. These questions, as well as the 
application of the derived
equations to other important problems of transport in solids such as the nuclear spin relaxation and diffusion, 
electro- and thermal conductivity, remain to be areas for further investigation.
%
%
%
\bibliography{kekuz}
%
%
%
%
%
%
%
%
%
\end{document}